\documentclass[review,twocolumn]{revtex4-1}
\usepackage{lineno,hyperref}
\usepackage{amsmath}
\usepackage{amssymb}
\usepackage{graphicx}
\usepackage{float}
\modulolinenumbers[10]

\begin{document}

\preprint{APS/123-QED}

\title{Anyon Black Holes}

\author{Maryam Aghaei Abchouyeh$^{1,2}$}
   \email{m.aghaei@ph.iut.ac.ir}
   \author{Behrouz Mirza$^{1}$   }
  \email{b.mirza@cc.iut.ac.ir}
   \author{Moein Karimi Takrami$^{1}$}
   \author{Younes Younesizadeh$^{1}$}

   \affiliation{$^{1}$ Department of Physics, Isfahan University of Technology, Isfahan 84156-83111, Iran}
   \affiliation{$^{2}$ Research Institute for Astronomy and Astrophysics of Maragha (RIAAM) $-$ Maragha, IRAN, P. O. Box: 55134 - 441}








\begin{abstract}
We propose a correspondence between an Anyon Van der Waals fluid and a (2+1) dimensional AdS black hole. Anyons are particles with intermediate statistics that interpolates between a Fermi-Dirac statistics and a Bose-Einstein one. A parameter $\alpha$ ($0<\alpha<1$) characterizes this intermediate statistics of Anyons. The equation of state for the Anyon Van der Waals fluid shows that  it has a quasi Fermi-Dirac statistics for $\alpha > \alpha _c$, but a quasi Bose-Einstein statistics for $\alpha< \alpha _c$. By defining a general form of the metric for the (2+1) dimensional AdS black hole and considering the temperature of the black hole to be equal with that of the Anyon Van der Waals fluid, we construct the exact form of the metric for a (2+1) dimensional AdS black hole. The thermodynamic properties of this black hole is consistent with those of the Anyon Van der Waals fluid.  For $\alpha< \alpha _c$, the solution exhibits a quasi Bose-Einstein statistics. For $\alpha > \alpha _c$ and a range of values of the cosmological constant, there is, however, no event horizon so there is no black hole solution. Thus, for these values of cosmological constants, the AdS Anyon Van der Waals black holes have only quasi Bose-Einstein statistics.


\end{abstract}

\maketitle

\section{\label{sec:level1}Introduction}
  The physics of the black holes has always been  one of the most interesting research areas since its appearance as a research field \cite{B,B1,HHH,HHH1,HHH2,H,BCH}. It is known that there is an analogy between an AdS black hole and the Van der Waals fluid \cite{KM}. The correspondence between the two is important because the thermodynamic behavior of black holes can be explained by that of the fluid; the conventional thermodynamic phase space (including temperature, entropy, and volume) can also be defined for an AdS black hole. This analogy will be more complete in an extended phase space. The past few years has witnessed an interest in the study of the cosmological constant ($\Lambda$) as a thermodynamic parameter in the first law of thermodynamics \cite{DST,KM,53,HT,HT1,T}. Although this assumption seems awkward, there are good reasons why $\Lambda$ should be considered in the first law of thermodynamics. First, including the cosmological constant $\Lambda$ in the first law of thermodynamics will make it consistent with Smarr relation\cite{53} and the variation of $\Lambda$ will satisfy the Smarr relation. Second, there exist theories that show physical constants such as $c$, $G$, $h$ and $\Lambda$ are not really constant and have variation with respect to the energy scale of the universe. \\
  \indent Once we introduce the cosmological constant as a thermodynamic parameter, we can define its conjugate variable. Since $\Lambda$ is proportional to the thermodynamic pressure (for $d$ dimensional space-time $P= -\dfrac{\Lambda}{8\pi}=\dfrac{(d-1)(d-2)}{16\pi{l^2}}$  using geometric units $G_N = \hslash = c = k_B = 1$)  , its conjugate must have volume dimension. This definition will give rise to an additional term, $P\delta V$, in the first law of thermodynamics and the mass of the black hole will be defined in terms of its enthalpy. In this extended phase space, one can write the equation of state for the AdS black hole and compare it with the equation of state for the Van der Waals fluid  that reads as follows:
  \begin{equation}
T=(P+\dfrac{a}{v^2})(v-b),
\label{2}
\end{equation}
where, $P$ is the thermodynamic pressure, $v$ is the specific volume of the fluid $v=\dfrac{V}{N}$, and $N$ is the degree of freedom ($V$ is the conjugate volume for $P$).  In Ref. \cite{VWBH}, the authors derived an exact form for the metric of an AdS black hole which has the same thermodynamics as the Van der Waals fluid. In this work, we construct an Anyon Van der Waals fluid  whose thermodynamics is exactly consistent with that of an AdS Anyon Van der Waals black hole. What they have in common is that the particles are considered to be either fermions or bosons which obey the Fermi-Dirac or the Bose-Einstein statistics, respectively. For a fluid of the latter type, there can be a Bose-Einstein condensation (For fermions there is no condensation  because of Pauli's exclusion principle). But in the (2+1) dimensional space time, we may have an intermediate statistics \cite{LMC,frank,MM,R,b}. The particles that have this intermediate statistics are called Anyons. It is notable that fermions and bosons are the two limits of the Anyons. Thus, we can use a real number  $\alpha$ ($0<\alpha <1$) to parameterize the intermediate statistics of the Anyons with $\alpha =0$  corresponding to bosons (the particles that can have the Bose-Einstein condensation), $\alpha =1$ corresponding to fermions (the particles which obey the Pauli exclusion principle), and  $0<\alpha <1$ parameterizing the intermediate statistics of the Anyons.
Here, we are going to construct a metric for a (2+1) dimensional black hole with statistics consistent with the intermediate statistics of the Anyon fluid. The results show that the Anyon Van der Waals fluid has a quasi Fermi-Dirac statistics for $\alpha_c<\alpha <1$ and that the AdS Anyon Van der Waals fluid has a quasi Bose-Einstein statistics for $0<\alpha <\alpha_c$. In the former case, however, there will be no black hole solution.  Thus, for $\alpha_c<\alpha <1$, it is not possible to describe an AdS Anyon Van der Waals black hole by means of an Anyon Van der Waals fluid. The interesting consequence of our work is  that AdS Anyon Van der Waals black holes can be expressed only for $0<\alpha <\alpha_c$ and that they have a quasi Bose-Einstein statistics.

The paper is organized as follows:\\
In Sec. $II$, we present a review of the AdS Van der Waals black hole and the equations for both its energy density and pressure.  In Sec. $III$, the equation of state for the Anyons is introduced. In Sec. $IV$, the exact form of the metric that is consistent with the AdS Anyon Van der Waals fluid is obtained, the equations of the energy density and the pressure of the black hole are derived, and the behavior of the energy density and the pressure are analyzed. It is interesting that there are black hole solutions that only correspond to the semi Bose Einstein statistics. We present the results and conclusions in Sec. $V$. 

\section{\label{sec:level2}Van der Waals Black hole}
 In \cite{VWBH}, the authors constructed a metric for a 3+1 dimensional AdS black hole that has a similar  thermodynamic behavior to that of the Van der Waals fluid and checked the validity of energy conditions for this black hole. This metric construction is based on the AdS Black hole similarity to Van der Waals fluid together with the assumption that the cosmological constant is a thermodynamic variable. In this extended phase space, the relation $P= -\dfrac{\Lambda}{8\pi}=\dfrac{3}{8\pi{l^2}}$ hold between the thermodynamic pressure and the cosmological constant $\Lambda$. By assuming this equation to be true, we should identify the conjugate variable for the pressure proportional to $\Lambda$; obviously, the natural choice is volume.  So, the equation for the mass of the black hole should be modified from $\delta M=T\delta S$ to:

  \begin{equation}
\label{mass}
\delta M=T\delta S+V\delta P +...,
\end{equation}
This is why, in an extended phase space for the AdS black hole, the mass of the black hole is related to its enthalpy \cite{Brain}. Using Eq. \eqref{mass}, one can see that the thermodynamic volume $V$ can be obtained from:
\begin{equation}
\label{volume}
V={\left(\frac{{\cal\partial} M}{\partial P}\right)}_{s,...}.
\end{equation}
Now if we have the metric of the black hole, we can write the equation of state for the black hole as $P=P(V,T)$ and compare it with that for the fluid. For simplicity, one can assume the metric to be:
\begin{align}
\label{metric}
ds^2&=-fdt^2+\dfrac{dr^2}{f}+r^2d\Omega ^{2}\\
f&=\dfrac{r^2}{l^2}-\dfrac{2M}{r}-h(r,P),\\
M&=\dfrac{4}{3}\pi r_+^3 P-\dfrac{r_+}{2}h(r_+,P)
\end{align}
where, M is the mass  of the black hole and $h(r,P)$ should be determined accordingly. We assume this metric to be a solution for the Einstein field equation $G_{\mu \nu}+\Lambda g_{\mu \nu}=8\pi T_{\mu \nu}$. The energy momentum tensor is defined in an orthonormal basis by $T^{\mu \nu}=\rho e^\mu _0e^\nu _0+\sum _i p_i e^\mu _ie^\mu _i$, where $\rho$ is the black hole energy density and $p$ is its pressure. So, the pressure and the energy density of the black hole can be calculated by using the metric in Eq. (\ref{metric}):
\begin{eqnarray}
\label{energy density}
\rho =-p_1=\dfrac{1-f-rf^{\prime}}{8\pi r^2}+P\\
p_2=p_3=\dfrac{rf^{\prime \prime} +2f^{\prime}}{16\pi r}-P,
\end{eqnarray}
with the prime denoting the derivative with respect to $r$.\\
One should define the function $f$ such that the equation of state for the black hole is consistent with that of the Van der Waals fluid. The specific volume and temperature of the black hole are defined as functions of the black hole horizon and the thermodynamic pressure,

\begin{eqnarray}
\label{V}
&v=\dfrac{k}{4\pi r_+^2}[\frac{4}{3}\pi r^3_+-\frac{r_+}{2}\dfrac{\partial h(r_+, P)}{\partial P}]\\
\label{T}
&T=\dfrac{f^{\prime}}{4\pi}=2r_+P-\dfrac{h(r_+, P)}{4\pi} -\dfrac{1}{4\pi}\dfrac{\partial h(r_+, P)}{\partial r_+},
\end{eqnarray}

\noindent where, for a $d$ space time dimension, $k=\frac{4(d-1)}{d-2}$, $v=k\frac{V}{N}$ and $N$ is proportional to the horizon area as $N=\frac{A}{{L} _{pl}^2}$ with $A=4\pi r^2$. Since we expect the equation of state for the AdS black hole to be consistent with that of the Van der Waals fluid, we should compare the equation of state obtained from Eqs. \eqref{V} and \eqref{T} with that of the Van der Waals fluid. The direct relation between the specific volume and pressure of Van der Waals fluid and the temperature of the black hole will be obtained by combining  Eqs. \eqref{V},\eqref{T}, and \eqref{2}:

\begin{equation}
\label{Van der Waals}
2r_+P-\dfrac{h}{4r_+\pi}-\dfrac{h^{\prime}}{4\pi}=\left( P+\frac{a}{v^2}\right)\left( v-b\right).
\end{equation}

Where prime denotes the derivative with respect to $r_+$. By setting $h(r,P)=A(r)-PB(r)$, one can find a solution for Eq. (\ref{Van der Waals}). This leads to an equation in the form of  $F_1(r)P+F_2(r)=0$ in which $F_1$ and $F_2$ are the functions of $A$ and $B$ and their derivatives. By setting $F_1(r)=0$ and $F_2(r)=0$ separately, one can derive the solution for $h(r,P)$; hence, we will have the solutions for the energy density and pressure of the black hole.\\
In this work, we are going to construct new types of black holes whose statistics completely matches that of the Anyon Van der Waals fluid. Anyons are (2+1) dimensional particles with a statistics that interpolates between Bose-Einstein and Fermi-Dirac statistics. As mentioned, the parameter $\alpha$ is used to identify these particles so that we expect the Anyons to obey the Pauli exclusion principle for $\alpha_c<\alpha <1$ and to have a Bose- Einstein condensation for $0<\alpha <\alpha _c$. \\

\section{\label{sec:level3}Equation of State for Anyons}
In the four dimensional space time, we have only Fermi-Dirac and Bose-Einstein statistics. But in the (2+1)  dimensional one, there can be particles with intermediate statistics. For these particles, the statistics interpolate between bosons and fermions. These particles with intermediate statistics are called Anyons. In the following, we will introduce the equation of states for Anyons. Generalizing the equation of state for bosons and fermions into the intermediate form will yield the equation of state for Anyons.\\
The number of quantum states for N bosons or fermions which occupy G state is :

\begin{equation}
\label{quantum number}
W_b=\dfrac{(G+N-1)!}{N!(G-1)!},\hspace{0.5cm}W_f=\dfrac{G!}{N!(G-N)!}.
\end{equation}

The fractional statistics may be expressed by the following generalized equation:

\begin{equation}
\label{F statistics}
W=\dfrac{[G+(N-1)(1-\alpha)]!}{N![G-\alpha N-(1-\alpha)]!},
\end{equation}

\noindent where, $\alpha=0$ corresponds to the bosons and $\alpha=1$ corresponds to the fermions while $0<\alpha <1$ represents the Anyons which  have intermediate statistics. By assuming the number of particles and the energy of the fluid to be constant, we will have:
\begin {eqnarray}
\label{number}
N&=\sum N_i\\
\label{energy}
E&=\sum \varepsilon _i N_i.
\end{eqnarray}

\noindent With $\varepsilon _i$ being the energy of each one of the $N_i$ particles, the partition function for the grand canonical ensemble will be given by:
\begin{equation}
\label{partition}
Z=\sum _{{N_i}} W({N_i})\large exp{(\sum _i N_i (\mu _i-\varepsilon _i)/T)},
\end{equation}
where, $\mu$ and $T$ are the chemical potential and the temperature, respectively.
 By some straightforward calculations and assuming that $E=PV$, we can obtain the equation of state for Anyons up to the second order of $\lambda$, as follows \cite{MM,aeq}:

\begin{equation}
\label{Anyons}
PV= NT\big[1+(2\alpha -1)N\lambda ^2/4V\big],
\end{equation}
where, $P$ is the thermodynamic pressure, $V$ the volume, and $\lambda =\sqrt{\dfrac{2\pi}{mT}}$, where $m$ is the mass of the fluid particle. Note that $ E = PV$ is the equation of state for a non-relativistic fluid. Thus, we don't expect the Bose-Einstein condensation for the fluid being considered. It was shown that the statistical interactions are attractive for $\alpha <1/2$ and repulsive for $\alpha >1/2$. For $\alpha <1/2$, the statistics of Anyons is similar to that of bosons and for $\alpha >1/2$, the statistics is similar to that of fermions. It may be noticed that we are using the Haldane fractional statistics in the 2+1 dimensional space-time. The Haldane fractional statistics can, however, be applied to higher dimensions where the fractional statistics is an effective one for collective particles. In 2+1 dimensions, the fractional statistics is among the intrinsic properties of any particle. It then follows that there would exist Anyons only in 2+1 dimensions. In the first step, we generalize the equation of state of the Anyons defined in Eq. (\ref{Anyons}) to the equation of state of an Anyon Van der Waals fluid which will have the following form:
 \begin{equation}
 \label{V F statistics}
 (P+\dfrac{a}{v^2})(Nv-Nb)= NT\big[1+(2\alpha -1)N\lambda ^2/4V\big],
 \end{equation}
 \noindent where, $a$ is a positive constant representing the attraction between the particles of the Van der Waals fluid and $b$ is the volume of a single molecule ($a=b=0$ corresponds to the ideal fluid). In this  case, we can find the value of $\alpha =\alpha _c=\dfrac{1}{2}$ using the calculation similar to that in \cite{MM}. $\alpha _c$ is the boundary between the quasi Bose-Einstein and the quasi Fermi-Dirac statistics of the fluid so that, for $0<\alpha <\alpha _c$,  the fluid has a quasi Bose-Einstein statistics, whereas for $\alpha _c<\alpha <1$, Pauli's exclusion principle has the leading role. So the fluid has a quasi Fermi-Dirac statistics. In the following Section, we are going to study an AdS black hole whose thermodynamics corresponds to Eq. (\ref{V F statistics}).
 
\section{\label{sec:level4}Anyon Van der Waals Black Hole}
 The thermodynamic properties of the AdS black holes have been studied extensively \cite{CCK,CCK1,CCK2,CCK3}. We will construct a metric for a (2+1) dimensional AdS black hole whose thermodynamic properties are similar to those of an Anyon Van der Waals fluid \cite{MM}. The general metric for a (2+1) dimensional black hole will be \cite{NTZ}:
\begin{align}
\label{A metric}
ds^2&=-fdt^2+\dfrac{dr^2}{f}+r^2d\phi ^{2}\\
\label{A metric 1}
f&=\dfrac{r^2}{l^2}-8M-h(r,P),\\
\label{A metric 2}
M&=\pi r_+^2P-\frac{h(r_+,P)}{8}
\end{align}
where, the unknown function $h(r,P)$ should be obtained in such a way that the thermodynamic properties of this black hole will be consistent with those of the Anyon Van der Waals fluid. Let us again assume this metric to be a solution for the Einstein field equation $G_{\mu \nu}+\Lambda g_{\mu \nu}=8\pi T_{\mu \nu}$ in (2+1) space-time and $T^{\mu \nu}=\rho e^\mu _0e^\nu _0+\sum _i p_i e^\mu _ie^\mu _i$, where using Eq.\eqref{A metric}, we have: 

\begin{equation}
\label{tetrad}
e^{\mu}_a=\begin{bmatrix}
\frac{1}{\sqrt{f}} & 0 & 0 \\
0 & \sqrt{f} & 0 \\
0 & 0 & \frac{1}{r}
\end{bmatrix}
\end{equation}

Using the energy momentum tensor already defined in Sec. II, the energy density and the pressure of the black hole are derived as follows:
\begin{eqnarray}
\label{rho}
&\rho =-p_1= -\frac{f^{\prime}}{16\pi r}+P\\
\label{pressure}
&p_2= -\frac{f^{\prime \prime}}{16\pi }-P,
\end{eqnarray}
and for the 2+1 dimensional space-time, $P=\dfrac{-\Lambda}{8\pi}$ \cite{mann2}. \\
Furthermore, using the Einstein field equation and the energy momentum tensor definition will yield the energy conditions as follows:
\begin{eqnarray}
Weak:&\qquad  \rho \geq 0\ ,\         \rho+{p}_i \geq 0\\
Strong:&\quad  \rho+{\sum}_i{ p}_i\geq 0\,  \qquad   \rho+{p}_i \geq 0\\
Dominant:& \qquad \rho \geq |{p}_i|
\end{eqnarray}
The validity of these energy conditions can be ensured by calculating the exact function for $h(r,P)$. Now, to calculate  the function $h(r,P)$ the temperature of the black hole and the temperature of the Anyon Van der Waals fluid are assumed to be equal. Using the Eqs. (\ref{A metric}-\ref{A metric 2}) and \eqref{volume} we will find the equations for the temperature and the volume of a 2+1 dimensional AdS black hole as:
 
\begin{eqnarray}
\label{VV}
&&v=\dfrac{4}{\pi r}(\pi r_+^2-\dfrac{1}{8}\dfrac{\partial h(r_+,P)}{\partial P})\\
\label{TT}
&&T=\dfrac{f'(r_+)}{4\pi}=4r_+P-\dfrac{1}{4\pi}\dfrac{\partial h(r_+,P)}{\partial r_+}
\end{eqnarray}

  We can, thus, find a (2+1) dimensional black hole whose thermodynamics is exactly similar to that of an Anyon Van der Waals fluid.
By assuming that $h(r,P)=A(r)-PB(r)$ and using the equality between Eq.\eqref{TT} and the temperature obtained form Eq. \eqref{V F statistics}, we will have:
 \begin{figure}[H]
\centering
\includegraphics[width=0.4\textwidth]{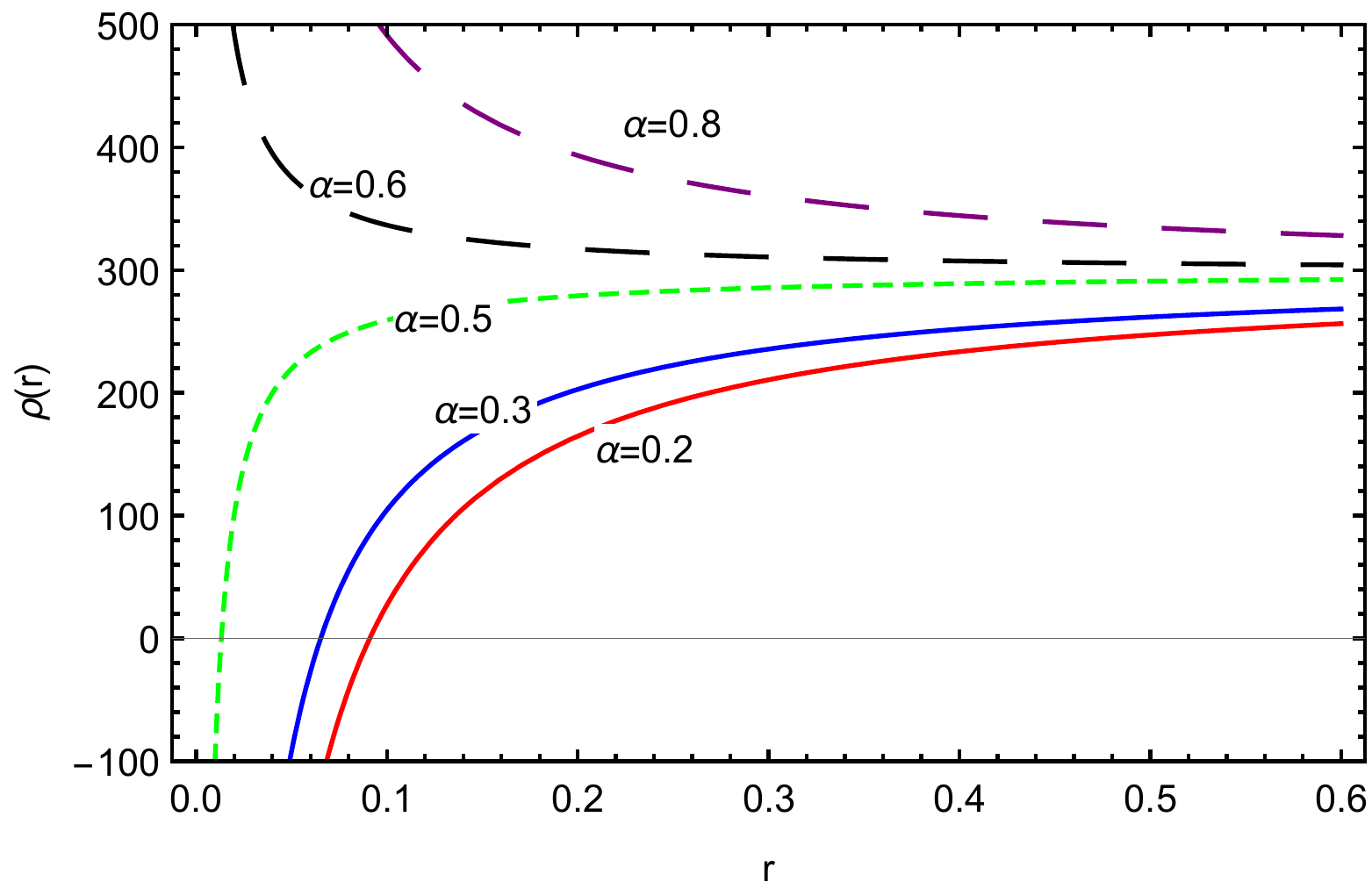}
\caption[]{\label{fig:figure 1b}\small{\it Energy density for the anyon Van der Waals black hole in the cases of $\alpha =0.2$ (red), $\alpha =0.3$ (blue), $\alpha =0.5$(dashed green), $\alpha =0.6$
(long dashed black) and $\alpha =0.8$ (long dashed perpule) for $M=100$, $a=\dfrac{1}{2\pi}$, $b=0.05$, $m=0.1$, $\epsilon=0.1$, $P=300$ and using the Eq. \eqref{Fp}. For $\alpha <\alpha _c=0.5$, the energy density of the black hole decreases  with decreasing values of  $\alpha$. Also, according to Eq. (\ref{V F statistics}) , there will be a quasi Bose-Einstein statistics for the Anyon Van der Waals fluid and
the energy density will decrease with decreasing values of $\alpha <\alpha _c=0.5$. Hence, the correspondence between these two cases.
For $\alpha >\alpha _c=0.5$, the energy density will increase with increasing values of $\alpha$ and the fluid will have a quasi Fermi-Dirac statistics. (For interpretation of the colors in the figure(s), the reader is referred to the web version of this article.)}}
\end{figure}
\begin{figure}[H]
\centering
\includegraphics[width=0.4\textwidth]{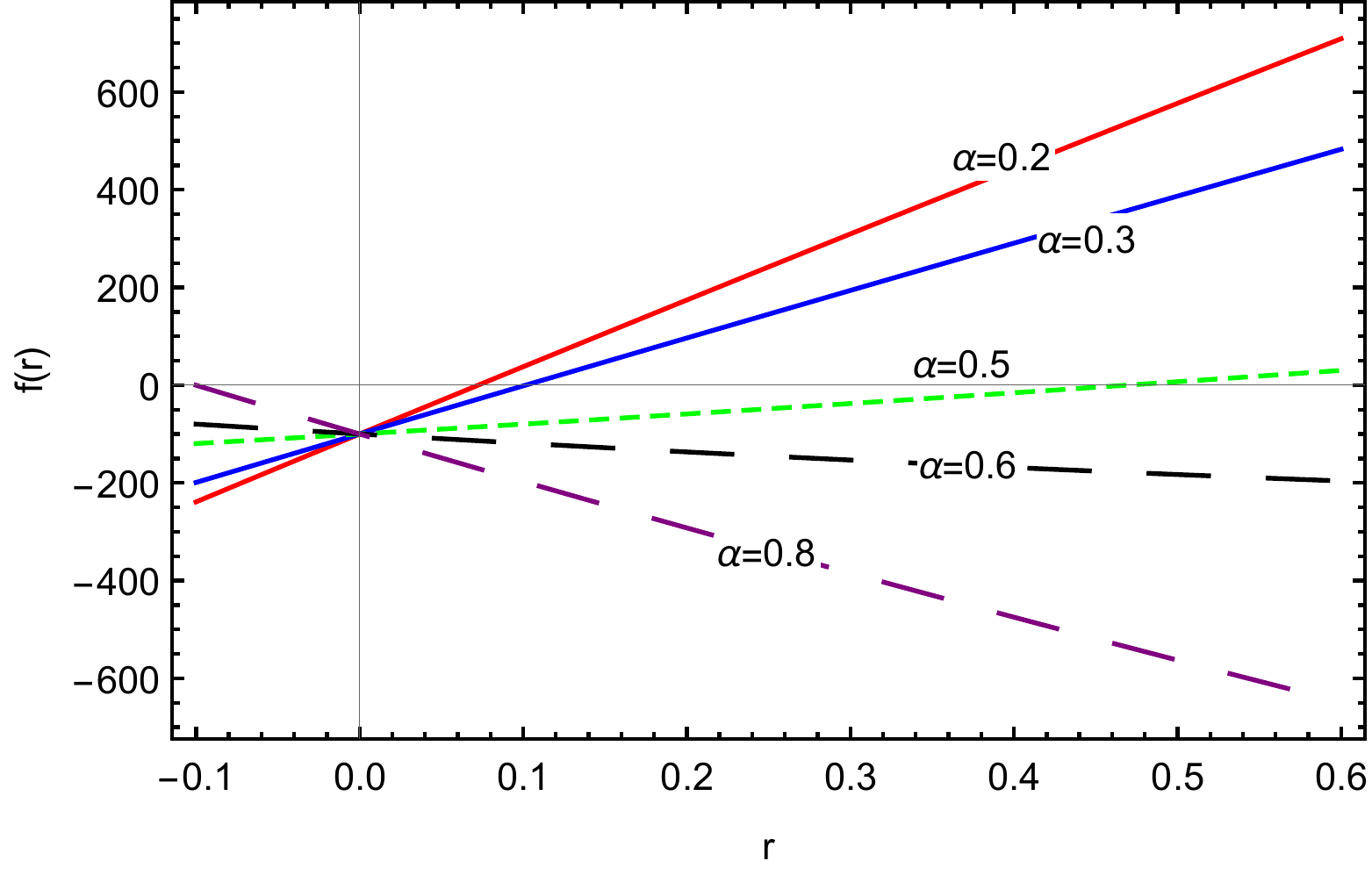}
\caption[]{\it{ The function $f(r)$ is plotted for $\alpha =0.2$ (red), $\alpha =0.3$ (blue), $\alpha =0.5$(dashed green), $\alpha =0.6$
(long dashed black) and $\alpha =0.8$ (long dashed perpule) for $M=100$, $a=\dfrac{1}{2\pi}$, $b=0.05$, $m=0.1$, $\epsilon=0.1$, $P=300$ and using the Eq. \eqref{Fp}. The horizon of the black hole is where the function $f(r)$ vanishes ($f(r)=0$). It is seen that there exists no horizon and, thereby, no black hole for $\alpha> \alpha _c=0.5$. For $\alpha \leq 0.5$, however, we can determine the position of the event horizon. $\alpha < 0.5$ corresponds to the era when the fluid has a quasi Bose-Einstein statistics. In this condition, the energy density of the black hole increases with $\alpha$.}}
\label{fig:figure 2b}
\end{figure}

\begin{figure}[H]
\centering
\includegraphics[width=0.4\textwidth]{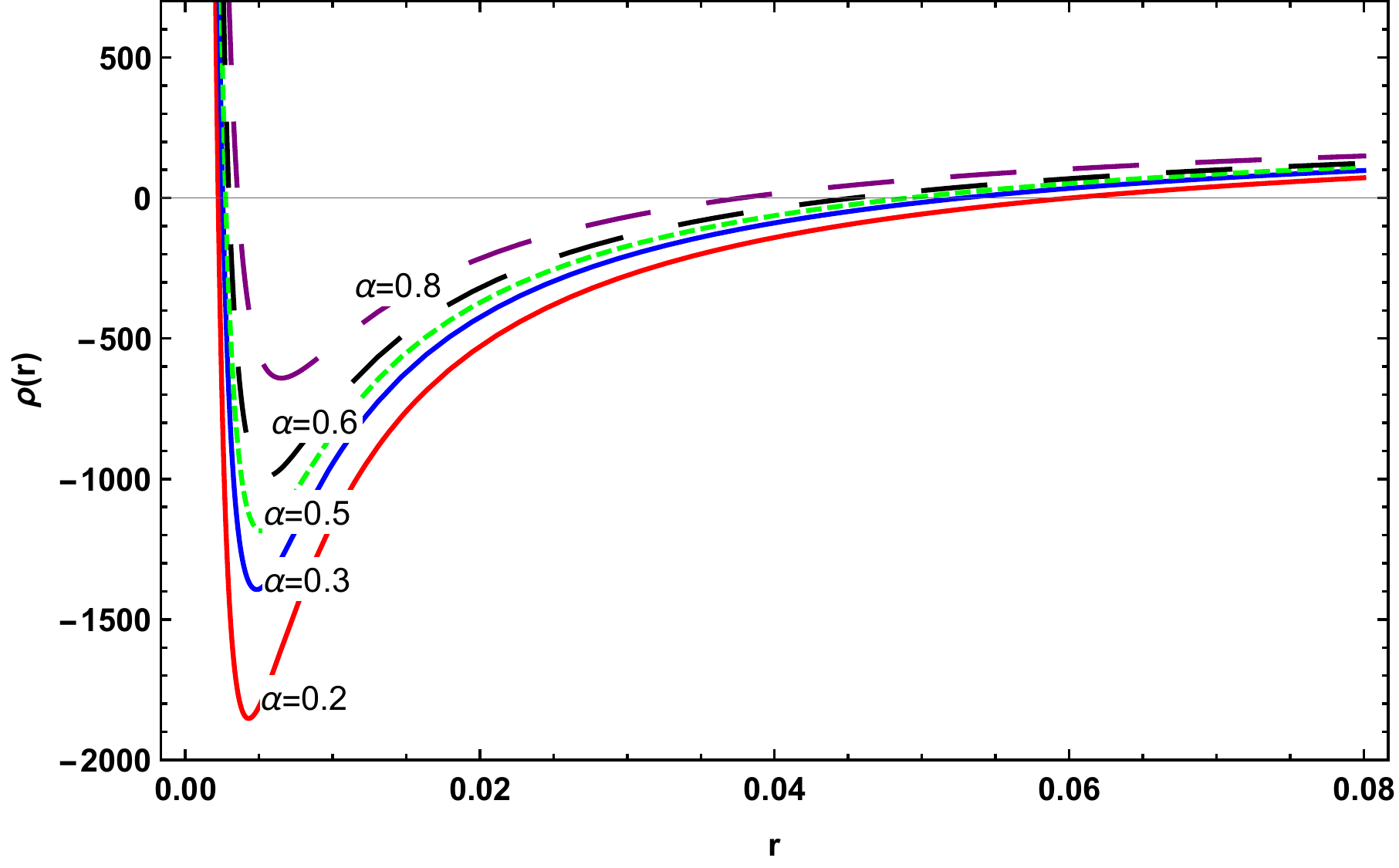}
\caption[]{\it Energy density of the Van der Waals black hole with intermediate statistics for 3+1 dimensions in the cases of $\alpha =0.2$ (Red), $\alpha =0.3$ (Blue), $\alpha =0.5$ (dashed green), $\alpha=0.6$ (long dashed black) and $\alpha=0.8$ (long dashed perpule) for $M=100$, $b=0.05$,$a=\dfrac{1}{2\pi}$, $m=0.1$ and $P=300$. The result remains the same for small values of $P$. } \label{fig:figure 4}
\end{figure}

\begin{figure}[H]
\centering
\includegraphics[width=0.4\textwidth]{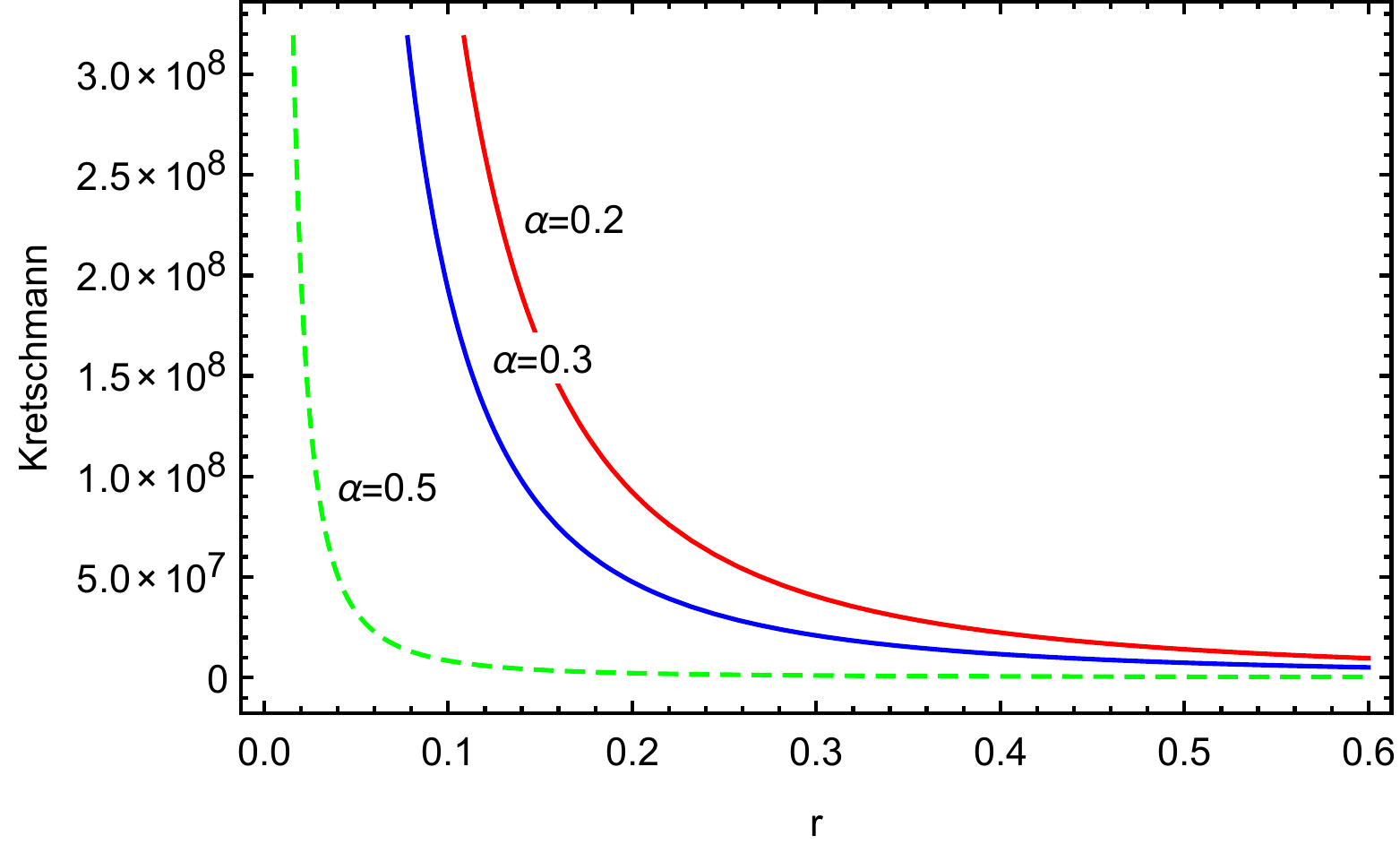}
\caption[]{\it The Kretschmann invariant for $\alpha =0.2$ (red), $\alpha =0.3$ (blue), $\alpha =0.5$ (dashed green) for $M=100$, $b=0.05$, $a=\dfrac{1}{2\pi}$, $m=0.1$, $P=300$ and $\epsilon =0.1$. } \label{fig:figure 3}
\end{figure}


\begin{equation}
\label{equality}
(P+\dfrac{a}{v^2})(v-b)-\dfrac{2\pi(2\alpha -1)}{4mv}=4 r P-\frac{A'(r)}{4 \pi }+\frac{P B'(r)}{4 \pi },
\end{equation}

\noindent where, $v=\frac{4 \left(\frac{B(r)}{8}+ \pi  r^2\right)}{\pi  r}$. The above equation is a differential one in the form $F_1(r)+PF_2(r)=0$  in which the functions $F_1$ and $F_2$ depend on the functions $A$ and $B$ and their first derivatives. The functions $F_1(r)$ and $F_2(r)$ should vanish independently and
we obtain two differential equations for $A(r)$ and $B(r)$, respectively. The solution for $B(r)$ is as follows:
\begin{eqnarray}
\label{A and B}
B(r)&=&4\pi br+c_1r^2
\end{eqnarray}
$c_1$ is an integration constant. At first we assume $c_1=-8\pi +\epsilon$, where $\epsilon$ is a small number ($\epsilon <1$). Now by replacing $B(r)=4\pi br+(-8\pi+\epsilon)r^2$ in the differential equation for $A(r)$ and expanding it up to the second order of $\epsilon$, we will have:

\begin{eqnarray}
\label{A DE}
\frac{1}{4} \left(-\frac{a m-2 \pi  \alpha +\pi }{b m}-\frac{A'(r)}{\pi }\right)+\nonumber\\
\frac{r^2 \epsilon ^2 (a m+\pi  (2 \alpha -1))}{64 \pi ^2 b^3 m}+\frac{\epsilon  (r-2 \alpha  r)}{16 b^2 m}=0
\end{eqnarray}

This equation can be solved for $A(r)$:

\begin{eqnarray}
\label{A}
 &&A(r)=\frac{1}{16 \pi  b^3 m}\Big (-16 \pi ^2 a b^2 m r+\frac{1}{3} a m r^3 \epsilon ^2+16 \pi ^3 (2 \alpha -1) b^2 r-\nonumber\\&&
 2 \pi ^2 (2 \alpha -1) b r^2 \epsilon +\frac{1}{3} \pi  (2 \alpha -1) r^3 \epsilon ^2\Big)
\end{eqnarray}

Now using Eq. \eqref{A metric 1}, the function $f(r)$ can be calculated as follows:

\begin{eqnarray}
\label{Fp}
&&f(r,P)=-M+P \left( 4 \pi  b r+r^2 \epsilon\right)-\frac{1}{16 \pi  b^3 m}\Big(-16 \pi ^2 a b^2 m r+\\
&&\frac{1}{3} a m r^3 \epsilon ^2+16 \pi ^3 (2 \alpha -1) b^2 r-2 \pi ^2 (2 \alpha -1) b r^2 \epsilon +\frac{1}{3} \pi  (2 \alpha -1) r^3 \epsilon ^2\Big) \nonumber
\end{eqnarray}

Thus in our model $\epsilon P$ is the effective cosmological constant. Now, the above solution may be replaced for $f(r,P)$  in Eq. \eqref{A metric} to solve the Einstein equations and obtain $\rho$ and $p$ in order to verify the energy conditions.
Using Eq. (\ref{rho}), the energy density of the black hole $\rho$ is depicted with respect to $r$ for different values of $\alpha$ in Fig. (\ref{fig:figure 1b}). As shown, for $\alpha <\alpha _c=\dfrac{1}{2}$, energy density decreases with decreasing  $\alpha$. From Eq. (\ref{V F statistics}), the quasi-Bose-Einstein statistics is expected for $\alpha <\alpha _c$. In our case, we have used $E=PV$, which is the equation of state for a non-relativistic fluid and the Bose-Einstein condensation is not expected for the fluid because the non-relativistic $2+1$ dimensional fluid cannot have this condensation \cite{mohammad zade}. \\

The black hole horizon is where the function $f(r)$ vanishes. Fig. (\ref{fig:figure 2b}) depicts the evolution of the function $f(r)$ with respect to $r$. It is interesting that for $\alpha >\alpha _c$ where the fluid has a Fermi-like behavior, there is no horizon, and, therefore, no black hole solution (Fig. (\ref{fig:figure 2b})). So, our method implies that for small values cosmological constant ($\epsilon<1$) black holes only exhibit a Bose-Like behavior in 2+1 dimensions. The important point is that, for the energy conditions to be satisfied in  2 + 1 dimension, the value of $\alpha$ in which the energy density is positive outside the horizon can be determined based on the values of other parameters. For the values of $P$, $a$ and $b$ here, energy density will be positive for $\alpha \leq 0.5$. These values of $\alpha$ also satisfy the Weak and Null energy conditions. For the BTZ metric the Kretschmann invariant is a constant, however for our solution it has a singularity at $r=0$. The Kretschmann invariants are depicted in Fig. (\ref{fig:figure 3}) for different values of $\alpha$. \\
For $\epsilon <1$ there is only black hole solution for $\alpha<0.5$; however, for other values of $\epsilon$ we may observe black holes with both Bose-like and Fermi-like statistics. It is interesting to add higher order corrections to Eq. \eqref{V F statistics} and  verify if there exists black hole solutions with Fermi-like statistics for $\epsilon >1$. We hope to study higher order corrections to Eq. \eqref{V F statistics} and explore corresponding black hole solutions in the future.\\
We may also study this statistical property of black holes in higher dimensions. In the $2+1$ dimension, the particles have intrinsic properties of Anyons, but in a $3+1$ dimension space-time, the particles may have an effective Haldane statistics as an effective intermediate one. So, the statistics  of the  collective particles may be checked. The energy density of the Black hole with an intermediate statistics in $3+1$ dimensions is depicted in Fig. (\ref{fig:figure 4}).  It may be noted that the fluid can be in the Bose-Einstein condensation state in 3+1 dimensions. As seen in Fig. (\ref{fig:figure 4}), in the $3+1$ dimension and for $\alpha<0.5$, the behavior of the fluid is Bose-like while it is Fermi-like for $\alpha>0.5$. The important difference in this case is that there are black hole solutions (horizons) in the 3+1 dimensions for all values of $\alpha$. 

Application of the Eq. (\ref{equality}) to the ideal fluid for which $a=b=0$  yields the same results and there would only exist  quasi Bosonic AdS Anyon Van der Waals Black holes.


\section{Conclusion}
In the three spatial dimensions, the metric of the space-time can be assumed to be in the form of Eq. (\ref{metric}). One can then obtain the exact form of the function $f(r)$ in such a way that the metric can describe the Van der Waals black hole \cite{VWBH}. This allows the functions of the energy density and pressure of these black holes to be calculated and their behavior to be studied.  For (2+1) dimensional AdS black holes, the general form of the metric should be considered in 2 spatial dimensions as in Eq. (\ref{A metric}), in which there can be particles with intermediate statistics. The two boundaries of the intermediate statistics are bosons ($\alpha =0$) and fermions ($\alpha =1$). Thus, $0<\alpha <1$ for the statistics interpolating between these two cases. The particles that have this intermediate statistics are called Anyons and they can exist in two spatial dimensions. Therefore, the (2+1) dimensional AdS black holes consist of an Anyon fluid. \\
We developed an exact form of the metric for the (2+1) dimensional AdS Anyon Black holes in which the Anyons obey the equation of state for the Van der Waals fluid. The equation of state was introduced in Eq. (\ref{V F statistics}) for an Anyon Van der Waals fluid. The metric in Eq. (\ref{A metric}) as well as the descriptions of the temperature and the volume of the black hole  were then used to obtain the exact form of the metric. Based on Eqs. (\ref{energy density}) and (\ref{pressure}), the exact form of $f(r)$ yields enough information not only to study of the behavior of the energy density and pressure of the black hole but also to investigate the existence of a horizon for the black hole. \\
Results show that for $M=100$, $m=0.1$, $a=\dfrac{1}{2\pi}$, $b=0.05$, $\epsilon <1$, $P=300$ and $0<\alpha < \alpha _{c}=0.5$, energy density decreases with decreasing value of $\alpha$ and we expect the fluid to have a quasi Bose-Einstein statistics of $\alpha$. For $ \alpha _{c}=0.5< \alpha <1$, we expect a Fermi-Dirac statistics for the fluid and  the energy density of the black holes to increase with increasing value of $\alpha$ (Fig. (\ref{fig:figure 1b})). As shown in Fig. (\ref{fig:figure 2b}), for $ \alpha _{c}=0.5<\alpha <1 $, there is no horizon for the Black hole assumed. This implies that there will be no real black hole  in this situation. Thus, it may be concluded that in the case studied here and for $M=100$, $a=\dfrac{1}{2\pi}$, $b=0.05$ there will be a (2+1) dimensional AdS Anyon Van der Waals black hole with a behavior similar to the Bose-Einstein statistics only for $\alpha < \alpha _{c}=0.5$.


\section{Acknowledgement}
This work has been supported financially by Research Institute for Astronomy and Astrophysics of Maragha (RIAAM) under research project No. 1/4717-72.

\end{document}